\documentclass[12pt,preprint]{aastex}
\newcommand{\myemail}{maggie@physics.mcgill.ca}
\newcommand{\psr}{PSR~B0540$-$69}

\newcommand{\rxte}{{\textit{RXTE}}}

\newcommand{\nudotdotdot}{{\ifmmode\stackrel{\bf \,...}{\textstyle \nu}\else$\stackrel{\,\...}{textstyle \nu}$\fi}}

\usepackage{graphicx}

\shorttitle{Phase-Coherent Timing of \psr}
\shortauthors{Livingstone et al.}

\begin{document}
\title{Long-term Phase-coherent X-ray Timing of \psr}

\author{Margaret A.~Livingstone \altaffilmark{1},
Victoria M.~Kaspi, Fotis P.~Gavriil}
\affil{Department of Physics, Rutherford Physics Building, 
McGill University, 3600 University Street, Montreal, Quebec,
H3A 2T8, Canada}

\altaffiltext{1}{\myemail}

\clearpage

\begin{abstract}
We present a new phase-coherent timing analysis for the young, energetic
pulsar \psr\ using 7.6\,yr of data from the \textit{Rossi X-Ray Timing
Explorer}. We measure the braking index, $n=2.140 \pm 0.009$, and discuss
our measurement in the context of other discordant values reported in the
literature. We present an improved source position from the phase-coherent
timing of the pulsar, to our knowledge, the first of its kind from X-ray pulsar timing.
In addition, we detect evidence for a glitch which has been previously
reported but later disputed. The glitch occured at MJD
$51335\pm12$ with $\Delta{\nu}/{\nu} = (1.4\pm0.2) \times 10^{-9}$
and $\Delta{\dot{\nu}}/{\dot{\nu}} = (1.33 \pm 0.02) \times 10^{-4}$. We
calculate that the glitch activity parameter for \psr\ is two orders of
magnitude smaller than that of the Crab pulsar which has otherwise 
very similar properties. This suggests that neutron stars of similar apparent 
ages, rotation properties and inferred dipolar B fields can have significantly
different internal properties. 
\end{abstract}

\keywords{pulsars: general---pulsars: individual (\objectname{\psr})---X-rays: stars}

\section{Introduction}
\label{sec:intro}

Young, rotation-powered pulsars spin down sufficiently rapidly to provide 
interesting tests of the simple model of pulsar spin down, given by
\begin{equation}
\dot{\nu} = -K{\nu}^n , 
\end{equation} 
where $\nu$ is the spin frequency, $\dot{\nu}$ is the frequency derivative,
and $K$ is a constant related to the magnetic field, moment of inertia, and
angle between the spin and magnetic axes. The braking index,
$n$, is equal to 3 for magnetic dipole radiation in a vacuum. A measurement
of the second frequency derivative, $\ddot{\nu}$, allows for the calculation
of $n$. Measurements of $n$ give crucial insight into the electrodynamics of
pulsars. Only five such measurements have been made 
\citep{lps93,lpgc96,dnb99,ckl+00,lkgm05}. Interestingly,
all of these values are less than 3 and span a relatively wide range:
$n=1.4\pm0.2$ to $n = 2.91 \pm 0.05$.  The physical implications of 
$n<3$ are still debated; possible explanations include the removal
of angular momentum by a pulsar wind \citep{mp89} and a
time-dependent magnetic field \citep{br88}.

Some pulsars, especially the youngest ones, exhibit significant
deviations from the standard spin-down model. Two types of deviations are
known, timing noise -- a low-frequency random process of unknown physical origin
superposed on the deterministic spin-down of the pulsar -- and glitches,
sudden increases in rotation rate likely caused by catastrophic unpinning of
superfluid vortices \citep[e.g.][]{ls90}.

One of the objects for which $n$ has been reported, $\sim 1500$\,yr old \psr,
has had several contradictory values quoted in the literature. The measured
values were obtained from timing observations made using a variety of instruments
operating from radio frequencies into the X-ray energy range and have produced values
that range from $n=1.81 \pm 0.07$ to $2.74\pm 0.10$ \citep{zmg+01, oh90}. The
differences among the measured values clearly signal contamination by one or 
more of timing noise, unidentified glitches, or even pulse counting errors. 

In this paper, we re-examine 4.6\,yr of data from the \textit{Rossi X-ray Timing
Explorer} (\rxte) reported on by \citet{zmg+01} and \citet{cmm03} who arrived at 
contradictory conclusions. We also extend the 
data set to a total of 7.6\,yr. We
present a measurement of $n$ for \psr\ with a partially phase-coherent
analysis, and confirm it by performing a fully phase coherent analysis.
We show that there is indeed evidence for 
a glitch near MJD $51335 \pm 12$ as reported by \citet{zmg+01}, but disputed
by \citet{cmm03}. However, we show
that the correct value of $n$ is $2.140\pm 0.009$, within $2\sigma$ of that
reported by \citet{cmm03}. 

\section{Observations}
\label{sec:obs}
Observations of \psr\ were obtained using the Proportional Counter Array 
\citep[PCA;][]{jsg+96} on board {\textit{RXTE}}. The PCA consists of an 
array of five collimated
xenon/methane multi-anode proportional counter units (PCUs) operating
in the 2~--~60~keV range, with a total effective area of approximately
$\rm{6500~cm^2}$ and a field of view of $\rm{\sim 1^o}$~FWHM.
We used 7.6\,yr of public archival \textit{RXTE}\ observations collected in
``GoodXenon'' mode, which records the arrival time
(with 1-$\mu$s resolution) and energy (256 channel resolution) of every
unrejected event. We used all layers of each PCU in the 2--18~keV
range. A similar energy range was used by both \citet{zmg+01} and 
\citet{cmm03}, as it maximizes the signal-to-noise ratio for this source.
Because \psr\ was most often not the primary target of \rxte\ in these
observations, integration times range from $\sim$1 to $>$10\,ks,
resulting in a variety of signal-to-noise ratios for individual pulse
profiles.

\par
The observations were reduced using standard FITS tools as well as specialized 
software developed independently. Data from the different PCUs
were merged and binned at 1/1024~ms resolution. The data were then
reduced to barycentric dynamical time (TDB) at the solar system barycenter
using, at first, a position determined by \textit{Chandra} observations \citep{kma+01} and the
JPL DE200 solar system ephemeris. Once a timing solution was determined, we were
able to fit for the position of the pulsar, as explained in \S \ref{sec:position}. We 
then re-barycentered the data at the
best-fit timing position. Each time series was folded with 32 phase bins using the 
ephemeris obtained by \citet{dnb99}, as this ephemeris produced
the highest quality pulse profiles. Resulting profiles were cross-correlated 
in the Fourier domain with a high signal-to-noise ratio template created by 
adding phase-aligned profiles from all observations. We implemented 
a Fourier domain filter by using the first 6 harmonics in the 
cross-correlation, appropriate for this broad pulse profile. The 
cross-correlation produces an average time of 
arrival (TOA) for each observation with typical uncertainty 0.75\,ms. TOAs were
fitted to a timing model (see \S \ref{sec:timing})
using the pulsar timing software package TEMPO 
\footnote{www.atnf.csiro.au/research/pulsar/tempo}. 

\section{Phase-Coherent Timing Analysis}
\label{sec:timing}

To determine spin parameters for \psr, we phase-connected 7.6\,yr
of timing data, that is, accounted for each turn of the
pulsar. This is achieved by fitting TOAs to a Taylor expansion of pulse
phase, $\phi$. At time, $t$, $\phi$ can be expressed as: 
\begin{equation}
\phi(t)=\phi(t_0) + \nu_0(t-t_0) + \frac{1}{2}{\dot{\nu}_0}(t-t_0)^2 +
        \frac{1}{6}{\ddot{\nu}_0}(t-t_0)^3 + {\ldots}  ,
\end{equation}
where the subscript 0 denotes a parameter evaluated at the reference
epoch, $t_0$. TOAs and initial parameters are input into TEMPO, 
which gives as output refined spin parameters and 
residuals. To determine an approximate input
ephemeris, we performed a periodogram analysis on 5 observations
from MJD 51197 - 51206. This provided 5 values for $\nu$, on which we
performed a least-squares fit, resulting in an initial estimate for
$\dot{\nu}$. We then applied the center value of $\nu$ and the fit 
value of $\dot{\nu}$ to the phase data and
`bootstrapped' a more accurate ephemeris which we extended over long time
spans.

We performed two timing analyses on the data. In \S \ref{sec:partially_coherent}, 
we describe our partially phase-coherent
analysis, which determined spin parameters while 
minimizing the effects of timing noise and
covariances arising from a polynomial fit to the data. In \S
\ref{sec:fully_coherent} we describe our fully
phase-coherent timing analysis, done to confirm our
partially coherent results and examine the timing noise apparent in the data. 
\S \ref{sec:position} describes a position fit performed using
phase-coherent timing. 

\subsection{Partially phase-coherent timing analysis}
\label{sec:partially_coherent}

The traditional method of phase-connecting all available data has resulted
in many important pulsar timing results owing to the accuracy available with
absolute pulse numbering. However, in cases where timing noise seriously
contaminates the data, many higher-order frequency derivatives are required
to satisfactorily fit the data by removing red noise. This invariably
introduces serious covariances between the fit spin parameters, without
entirely removing the contamination from the timing noise. 

A partially 
phase-coherent analysis is less sensitive to contamination from
timing noise. In this method, we phase-connect subsets of the data, where the length 
of each subset is determined by it being 
the longest that requires only $\nu$ and $\dot{\nu}$ be fit while having
residuals consistent with white noise. In this way, the 7.6\,yr of data 
were divided into 22 subsets. The measured values of 
$\dot{\nu}$ in these subsets are shown in Figure \ref{fig:nudot}. A
discontinuity can
clearly be seen in this plot. We interpret it as a glitch that occured at
MJD 51342$\pm$24, where the epoch is taken to be the midpoint between 
the two neighbouring measurements, and the uncertainty is the range between
these values. In \S \ref{sec:fully_coherent} we confirm this interpretation
using the more traditional fully coherent fit. A more accurate glitch epoch
is impossible to ascertain from the partially coherent method due to sparse
sampling and the small magnitude of the glitch. 

A value for $\ddot{\nu}$ obtained from the slope of
the best-fit line to all of the data will clearly be different 
from that obtained in two separate fits of the pre- and post-glitch data. The pre-glitch 
value of $\ddot{\nu}$ implies $n=2.135 \pm 0.016$, while the
post-glitch value yields $n=2.144 \pm 0.007$, where the uncertainties are
obtained by a bootstrap analysis which gives accurate uncertainties in cases
where the formal uncertainties are thought to underestimate the true values,
i.e. in the presence of timing noise \citep{efr79}. The two values of 
$n$ are in agreement within the quoted
uncertainties; thus we quote the average of the pre- and post-glitch values,
$n = 2.140 \pm 0.009$.

Figure \ref{fig:nudot_resid} shows the measurements of $\dot{\nu}$
with the pre-glitch slope fitted out. This plot clearly shows the change in $\dot{\nu}$ at
the epoch of the glitch. The change in $\Delta{\dot{\nu}} / \dot{\nu}$ is $\sim
(1.5 \pm 0.1) \times 10^{-4}$. The significant trend in the post-glitch data in Figure
\ref{fig:nudot_resid} can be readily explained by timing noise.

\begin{figure}
\plotone{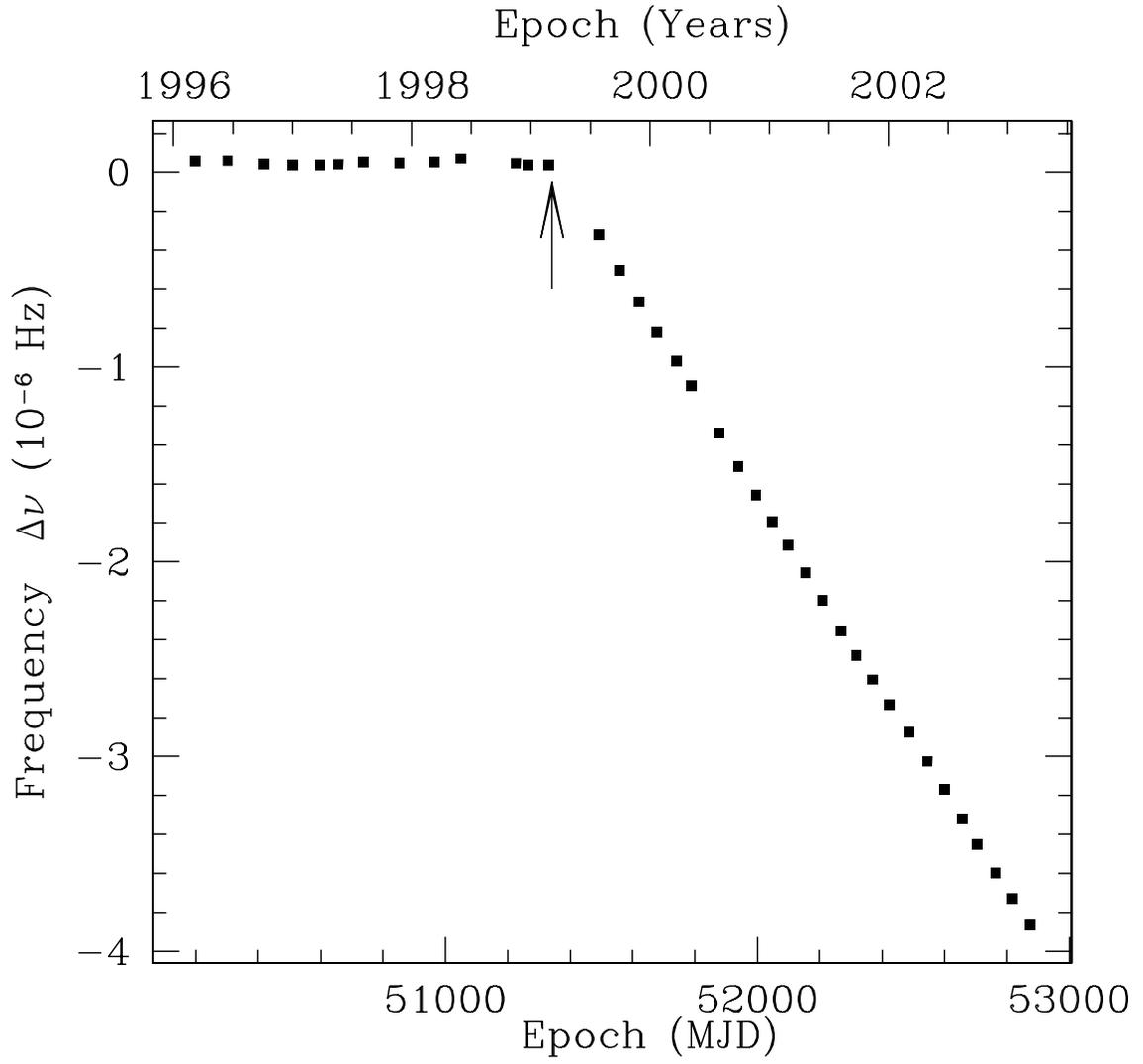}
\figcaption[f3.eps]{Measurements of $\nu$, with the pre-glitch trend
removed. The arrow marks the glitch epoch, MJD 51342. Uncertainties
are smaller than the points. 
\label{fig:freq_resid}}
\end{figure}

By further subdividing the data, we measured 38 values for $\nu$.
Fitting the pre-glitch trend ($\dot{\nu}$ and $\ddot{\nu}$) and removing it
from the data shows the influence of the glitch on $\nu$ as presented in 
Figure \ref{fig:freq_resid}. No jump in $\nu$ can be seen at the glitch epoch 
owing to the small magnitude of the glitch (as shown in \S \ref{sec:fully_coherent}). 
In fact, $\Delta{\nu}$ is on the same order as the uncertainty in the
frequency measurements, rendering the partially phase-coherent method
unsuitable for measuring a meaningful $\Delta{\nu}/\nu$ in
this case. Spin and glitch parameters obtained in the partially coherent 
analysis are given in Table 1.

\subsection{Fully Phase-Coherent Timing Analysis}
\label{sec:fully_coherent}
\begin{figure}
\plotone{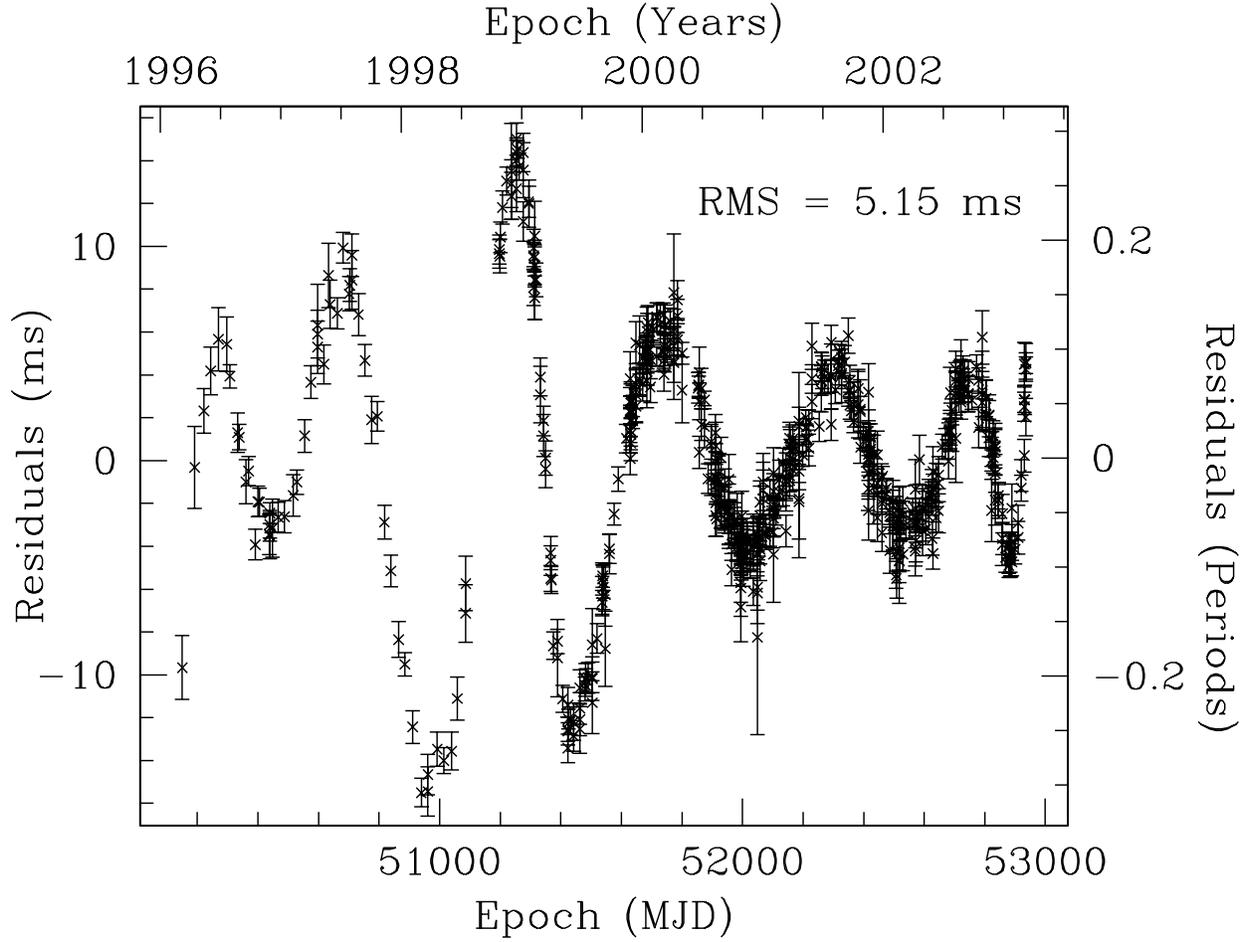}
\figcaption[f4.eps]{Residuals for \psr\ with $\nu$ and 11
frequency derivatives -- and no glitch -- fitted. This was the minimum
number of derivatives needed to connect the data, i.e. yield residuals
having amplitude less than 0.5~P. \label{fig:resids_noglitch}}
\end{figure}

To confirm the result obtained with the partially coherent timing analysis
and obtain more precise glitch parameters,
we phase connected all 7.6\,yr of timing data. We first connected the data
without allowing for a glitch in order to see if one was required by the 
fit. No phase jump was 
apparent at the epoch of the reported glitch, MJD 51325, 
however, we found that the timing residuals grew very significantly
when connecting over this epoch. In order
to phase connect the full data set with frequency derivatives and no glitch,
$\nu$ and 11 frequency derivatives were required, as shown in Figure
\ref{fig:resids_noglitch}, resulting in $n=1.968\pm0.004$. 

We then allowed for
a glitch near the epoch of that reported by
\citet{zmg+01}. In this case, the data can be fit well with 3 glitch
parameters (epoch, $\Delta{\nu}$, and $\Delta{\dot{\nu}}$)
and 3 spin parameters ($\nu$, $\dot{\nu}$, and $\ddot{\nu}$). Residuals for this
fit are shown in the top panel of Figure \ref{fig:resids_glitch}.  
The best fit glitch parameters, as determined with TEMPO, gives a
glitch epoch of MJD $51335 \pm 12$, in agreement with the partially
phase-coherent analysis.

We found $\Delta{\nu}/{\nu} = (1.4\pm0.2) \times 10^{-9}$, which was undetectable in
the partially coherent analysis, and 2.5$\sigma$ from the value reported by
\citet{zmg+01}. We also found
$\Delta{\dot{\nu}}/{\dot{\nu}} = (1.33 \pm 0.02) \times 10^{-4}$, in
agreement with the partially coherent analysis, but 10$\sigma$
larger than that reported by \citet{zmg+01}, $\Delta{\dot{\nu}}/{\dot{\nu}}
= (0.85 \pm 0.05) \times 10^{-4}$.

To conclusively show that the model with the glitch is a better description of the data,
we fitted additional frequency derivatives until the same number of
parameters was included in each fit (i.e. 12 parameters). The bottom panel
of Figure \ref{fig:resids_glitch} shows the results of this fit. Note that the RMS residual
is a factor of 5.15 smaller for the model with the glitch, corresponding to a
reduced ${\chi}^2$ of 1.86 for 545 degrees of freedom, compared to a reduced
${\chi}^2$ of 49.4 for 545 degrees of freedom for the fit with no glitch. This
confirms the conclusion of our partially coherent analysis, namely that a glitch occured
as reported by \citet{zmg+01}.

The phase-connected solution with the minimum number of derivatives  
(i.e. the number required to obtain phase-connection, in this case, 2), and the
glitch fitted, gives a value of
 $n=2.10657 \pm 0.00001$. However, since there is clearly low-frequency noise contaminating
the data (top panel of Figure \ref{fig:resids_glitch}), the formal uncertainty greatly
underestimates the true uncertainty on $n$. Spin parameters vary when higher order 
derivatives are fitted due both to timing noise and
covariances between parameters \citep{lkgm05}. Thus, we estimate the upper limit on
the uncertainty in $n$, albeit not rigorously, from the
variation in the measured value as higher order derivatives are added
to the fit. This method implies an uncertainty of $2.11 \pm 0.06$, in
agreement with the value obtained in the partially phase-coherent analysis. All spin
and glitch parameters from the fully phase-coherent analysis are given in
Table 1. Uncertainties given for parameters are the formal uncertainites, except
in the case of $n$. 

\subsection{Timing Position}
\label{sec:position}
\begin{figure}
\plotone{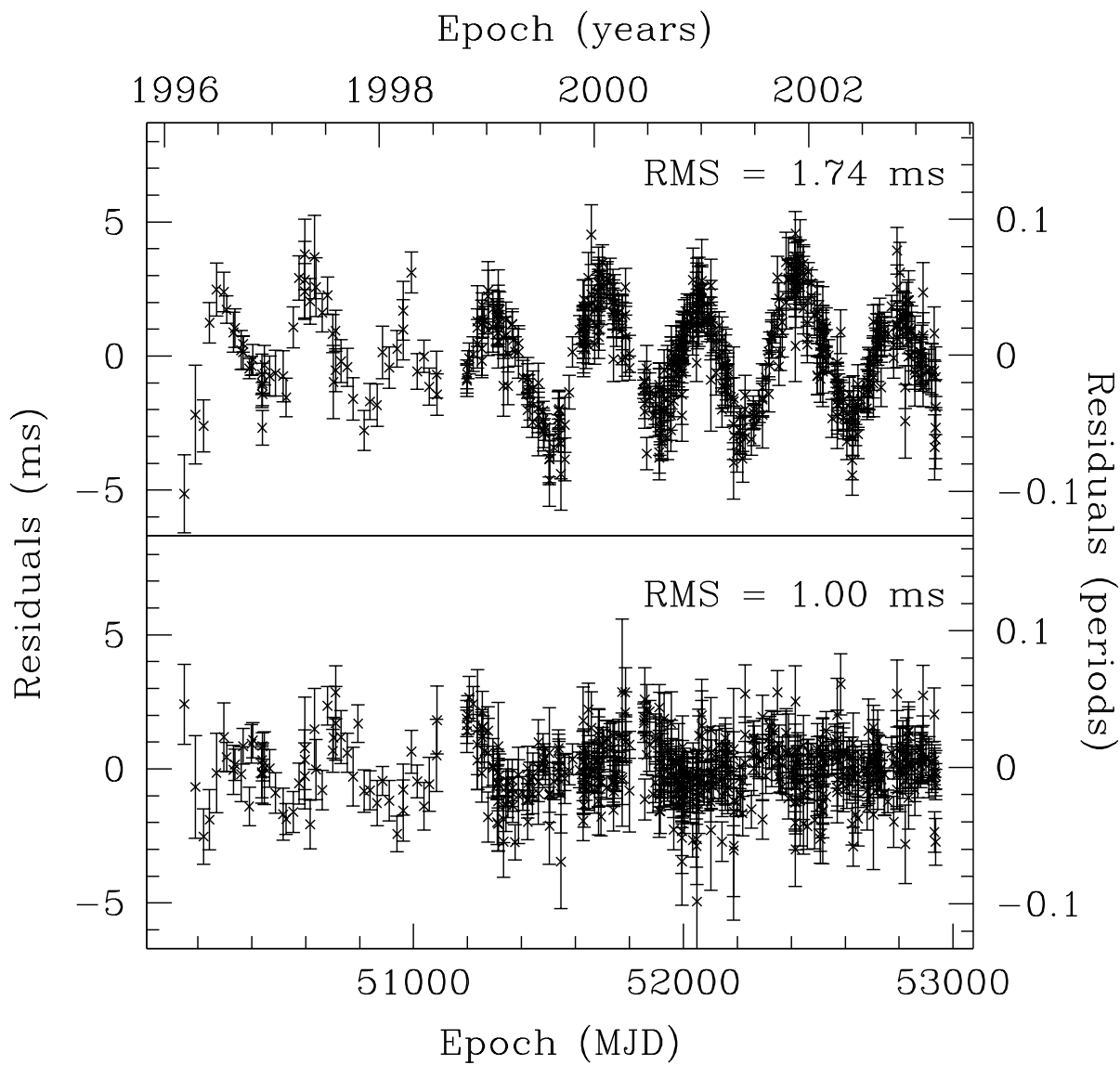}
\figcaption[f6.eps]{RMS residuals for the \textit{Chandra} position (top) and the
best-fit position of \psr\ (bottom). Both sets of residuals have had $\nu$
and 8 frequency derivatives fitted; removing higher order derivatives does
not improve either fit significantly. \label{fig:pos}}
\end{figure}

Our initial timing analysis was first performed 
using the pulsar position determined in a \textit{Chandra} observation
\citep{kma+01}. The reported 1$\sigma$ uncertainty was 0.7\arcsec\ on the
\textit{Chandra} position. Holding this position fixed, the post-fit 
residuals for the fully phase-coherent analysis showed
periodic behaviour, with a period of 1 year and amplitude $\sim5$\,ms, shown
in the top panel of Figure \ref{fig:pos} (which also has fitted the glitch
and 8 frequency derivatives as in Figure \ref{fig:resids_glitch}). This 
suggested a small but highly significant position hence barycentering
error. We found that the periodicity in the residuals changes negligibly
when higher order derivatives are fitted, implying that a polynomial does
not describe the trend in the data. We performed a grid search for position
by re-barycentering the data and
calculating TOAs for each new position. We then phase-connected each set of
new TOAs and calculated post-fit residuals. We found that the position that
minimizes the RMS residuals is RA=05$^{\rm{h}} 40^{\rm{m}}
11\fs16\pm0\fs04$ and
Dec$=-69^{\circ} 19' 53\farcs9\pm0\farcs2$, $1\farcs3$ from the \textit{Chandra}
position. The phase-coherent
RMS residuals fall by a factor of $\sim 2$ and the reduced $\chi^2$ is smaller
by a factor of 4 when the fitted position is used. More importantly, the periodic
trend in the data is removed by this fit, shown in the bottom panel of Figure
\ref{fig:pos}. Remaining systematics are likely due to
timing noise. All quoted parameters for the phase-coherent and partially 
phase-coherent analysis are calculated using
the new position. However, none of the spin parameters is significantly
different from what is obtained using the \textit{Chandra} position. To our
knowledge, this is the first time phase-coherent X-ray timing has been used
to try to improve a source position. If correct, we have effectively demonstrated superior
source localization using the non-imaging, 1$^{\circ}$-FOV PCA compared with
that attainable with \textit{Chandra}. However, it is possible that the
source timing noise is merely mimicking a position error; continued timing
will confirm or refute this result.

\section{Discussion}
\label{sec:discussion}

\subsection{The Braking Index}
By performing both a phase-coherent and a partially phase-coherent analysis
on 7.6\,yr of timing data for \psr, we have found $n=2.140 \pm 0.009$. 
This value, stationary across the glitch epoch, is similar to that reported
by \citet{dnb99}, $n=2.080  \pm 0.003$ based on five years of GINGA data, and
1.7$\sigma$ from that obtained by \citet{cmm03}, $n = 2.125 \pm 0.001$,
based on 4.6\,yr of \rxte data. \citet{cmm03} however, reported no glitch
and their uncertainty on $n$ does not
account for the effects of timing noise. The reason for the good agreement 
between our measured values is that
despite the fact that they report no glitch, the value given for $n$ is
obtained from phase-coherent fits to the data before and after the glitch
reported by \citet{zmg+01}, instead of a fit to all 4.6\,yr as would be
appropriate if no glitch had occurred. We phase connect the same 4.6\,yr
of data they used without a glitch and find $n=1.73 \pm 0.03$,
significantly different from the value obtained not considering the glitch,
further implying the occurance of a glitch.

Our value of $n$ is significantly larger than that reported by
\citet{zmg+01}, $n=1.81 \pm 0.07$. We find that if we phase-connect only
the same 300-day data set they used, we obtain a value $n = 1.82 \pm
0.01$, in agreement with their result. Thus their measurement was clearly
contaminated by timing noise and/or glitch recovery, and affected by the
relatively small time baseline used to measured $n$.

\par
The value of $n=2.140 \pm 0.009$ is significantly less than that for simple
magnetic dipole radiation, $n=3$. In fact, all measurements of $n$ thus
far are less than 3. Several explanations for this have been
proposed. Loss of angular momentum owing to a particle wind (which
young pulsars are well known to have) could account for this \citep{mp89}. 
Another explanation is that the neutron star loses additional rotational
energy by torquing a disk of supernova fallback material \citep{mph01}. Another
explanation is a time-varying magnetic field \citep{br88}. The presence of a
time-varying magnetic field can be determined by measuring a value of the
third frequency derivative larger or smaller than that predicted by the
spin-down model. The simple spin-down
model given in Equation 1 assumes a constant value of $K$, which is only
valid if the magnetic field, moment of inertia, and angle between the
spin and magnetic axes are all constant. With present data, the third
frequency derivative is not measurable for \psr\ because of contamination
from timing noise. It is possible that, as observations continue, this
parameter will be measurable. For both the Crab pulsar and PSR B1509$-$58,
values of $\nudotdotdot$ have been measured and are consistent with the
spin-down law \citep{lps93,lkgm05}. 

\subsection{Glitches and Timing Noise}

\begin{figure}
\plotone{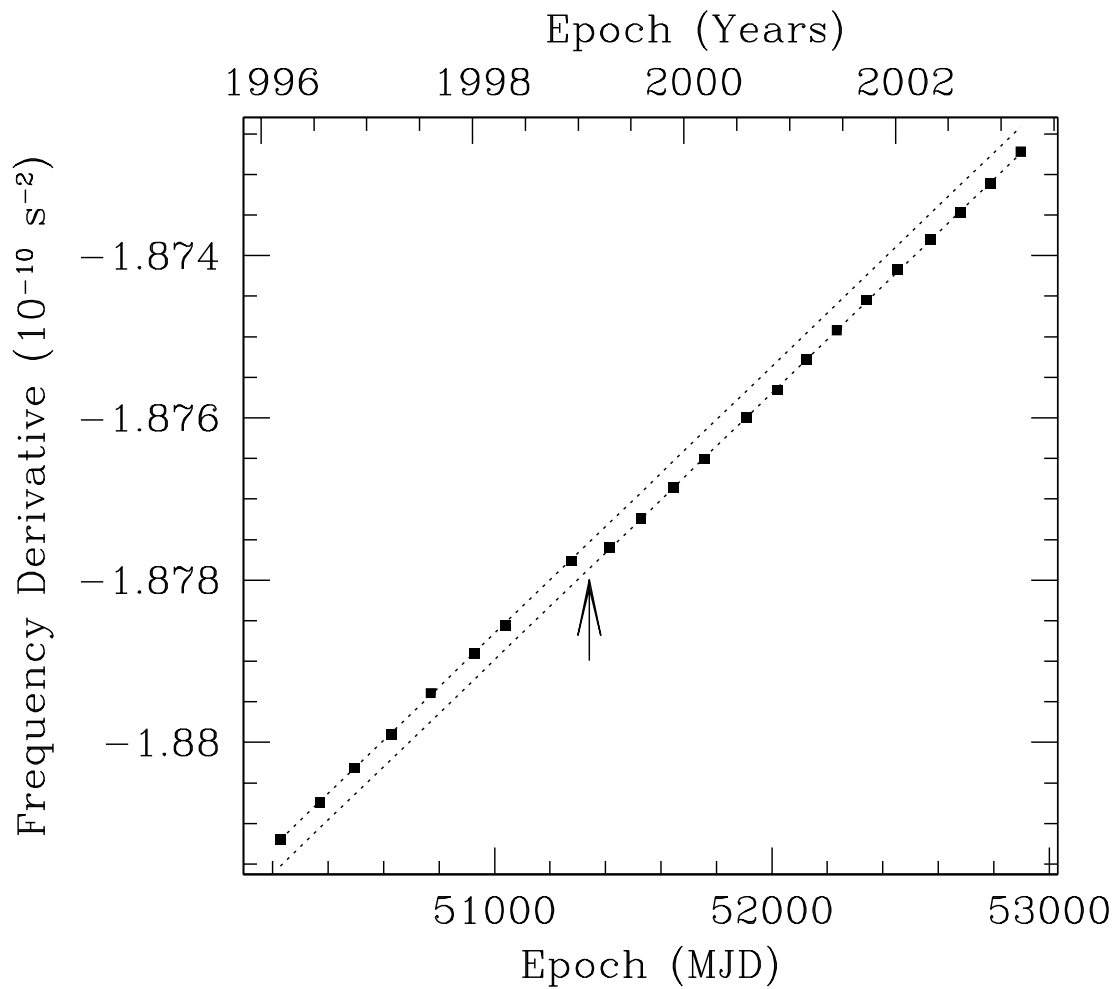}
\figcaption[f1.eps]{Measurements of $\dot{\nu}$, the slope is
$\ddot{\nu}$. The glitch occuring near
MJD 51342 is shown with an arrow. The pre-glitch 
value is $\ddot{\nu}=(3.81\pm 0.03) \times 10^{-21}$\,s$^{-3}$ implying 
$n=2.315\pm0.016$, while the post-glitch value is $\ddot{\nu}=(3.81\pm 0.01)
\times 10^{-21}$\,s$^{-3}$ implying $n=2.144\pm 0.007$. The average of
pre- and post-glitch $n$ is $2.140 \pm 0.009$. Measurement uncertainties 
are smaller than the points and are omitted. \label{fig:nudot}}
\end{figure}

\begin{figure}
\plotone{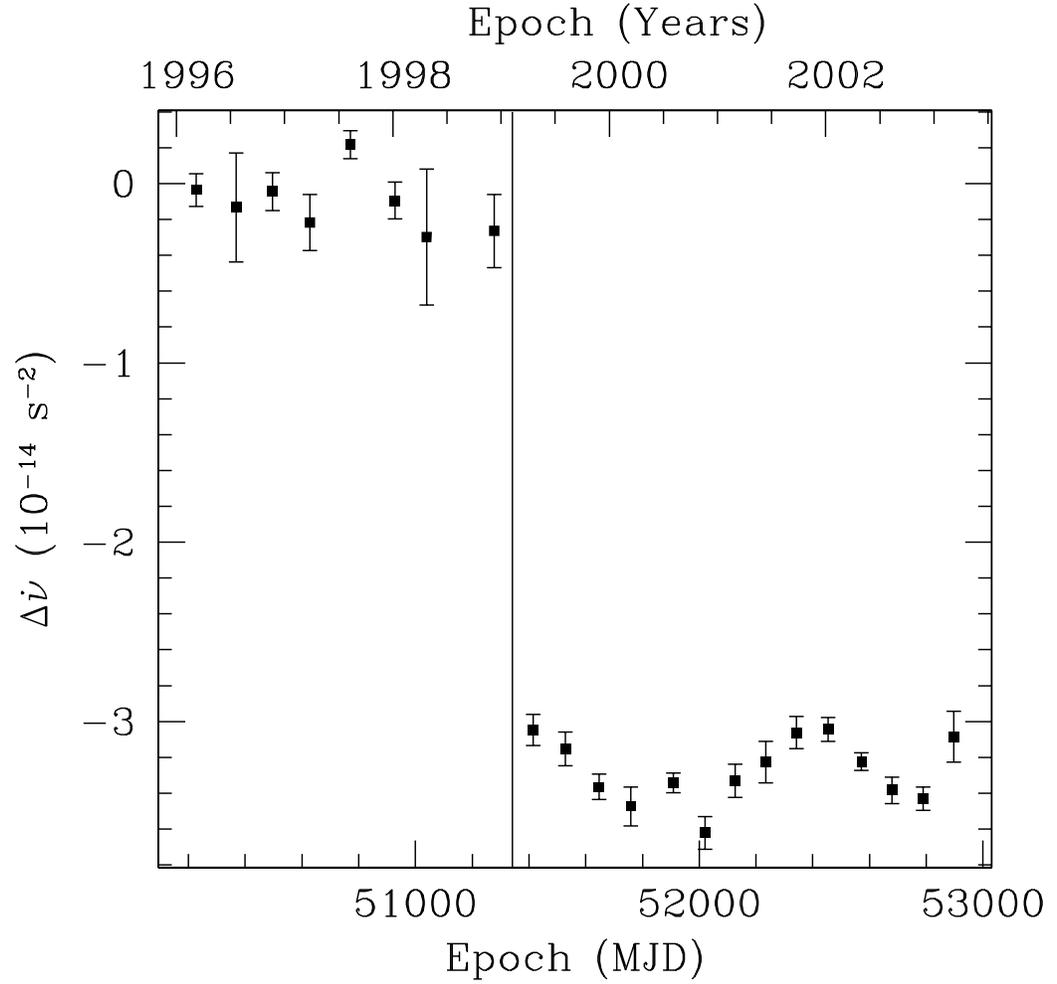}
\figcaption[f2.eps]{Measurements of $\dot{\nu}$ with the
pre-glitch slope (ie. $\ddot{\nu}$) subtracted. The value of $\Delta{\dot{\nu}}/{\dot{\nu}} \sim (1.5 \pm
0.1) \times 10^{-4}$ is in agreement with the value obtained from a 
phase-coherent analysis. The vertical line indicates the approximate glitch epoch, 
MJD 51342. The trend in the post-glitch data is attributable to timing noise.
\label{fig:nudot_resid}}
\end{figure}

\begin{figure}
\plotone{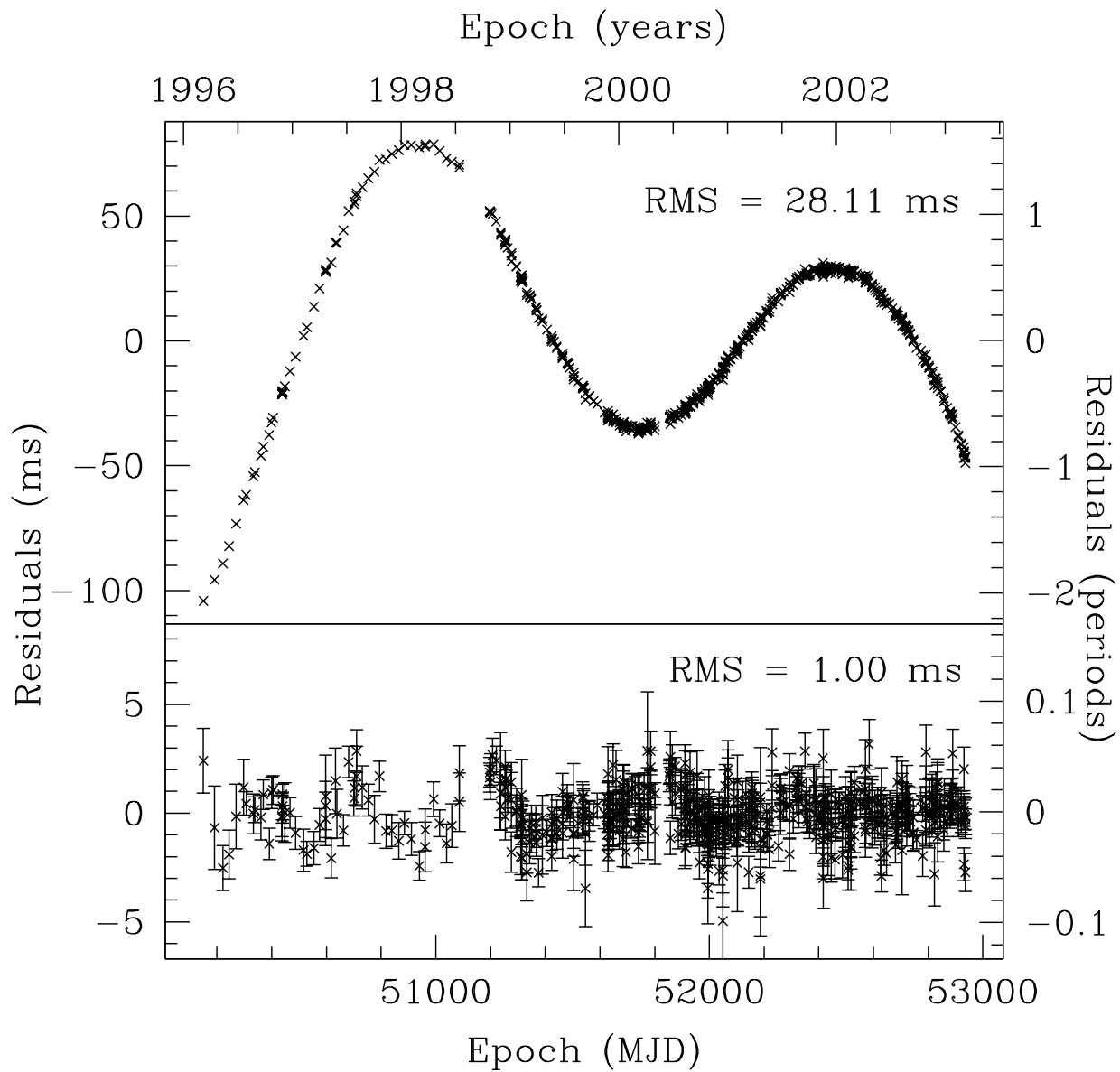}
\figcaption[f5.eps]{Timing residuals for \psr. Top panel
shows residuals with three glitch parameters and two frequency derivatives fitted. This fit
yields a value of $n=2.11 \pm 0.06$. Bottom panel shows residuals with 12
parameters total fitted, for comparison with Figure~\ref{fig:resids_noglitch}.
\label{fig:resids_glitch}}
\end{figure}

We found that the best description of the 7.6\,yr of \rxte\ data include a
glitch at MJD $51335 \pm 12$. Our partially phase-coherent analysis shows a
clear discontinuity in $\dot{\nu}$ (Figures \ref{fig:nudot} and
\ref{fig:nudot_resid}). The fully phase coherent analysis confirms this
result. This glitch was previously reported by \citet{zmg+01},
at MJD $51325 \pm 45$ and later refuted by \citet{cmm03}, who claimed that the
data could be described by fitting higher order derivatives, that is, timing
noise. The detected glitch has a small magnitude in both $\Delta{\nu}/{\nu}$
and $\Delta{\dot{\nu}}/{\dot{\nu}}$ which \citet{cmm03} cite as evidence that
no glitch occurred. However, fitting derivatives to the data shows that
the residuals increase dramatically when fitting over the epoch of the reported 
glitch. This is exactly what is expected from a glitch, and is not expected 
from timing noise, which increases residuals gradually as observations are
added. In addition, the partially coherent analysis shows a clear
discontinuity in $\dot{\nu}$ near (within uncertainties) the epoch of the
glitch found by the phase-coherent analysis (Figure \ref{fig:resids_glitch}). 
Sparse observations of this source previous to the
\rxte\ timing program most likely would not have been able to distinguish
a small glitch from timing noise. Undetected glitches of the same magnitude
as the one reported here are a likely explanation of the wide range of 
reported values of $n$. 

Although the magnitude of the glitch is small, it is similar to the smallest
glitches experienced by the Crab pulsar, which is close in
age, spin-down luminosity ($\dot{E} ={-4{\pi}^2I{\nu}{\dot{\nu}}}$), and
dipole magnetic field ($B_0 \sim 3.3 \times 10^{19} (P\dot{P})^{1/2}$) to
\psr. In fact, \psr\ is often referred to 
as the ``Crab Twin'' due to these similarities. The Crab pulsar has
glitched with parameters as small as $\Delta{\nu}/{\nu} \sim 2 \times
10^{-9}$ and $\Delta{\dot{\nu}}/{\dot{\nu}} \sim 10^{-5}$ \citep{lps93}.
However, the Crab pulsar also experiences glitches on the order of
$\Delta{\nu}/{\nu} \sim 10^{-8}$ with values of 
$\Delta{\dot{\nu}}/{\dot{\nu}}$ as large as $4 \times 10^{-4}$ \citep{wwm+01}. 
None of these larger magnitude glitches has been observed in \psr.

The glitch activity parameter, A$_g$, is defined as the cumulative change in
frequency from all glitches over the observation length, A$_g = \Sigma(
\Delta{\nu}/{\nu})/{\Delta{t}}$ \citep{ml90}. The Crab pulsar has
been observed to glitch 14 times between 1969 and 2000 and has
a glitch activity parameter A$_g \simeq 0.1 \times 10^{-7}$\,yr$^{-1}$
\citep[e.g.][]{sl96}. PSR~J1119$-$6127, aged $\sim1600$\,yr with a measured
braking index, $n=2.91\pm0.05$ has been observed to glitch once implying
A$_g \simeq0.02\times 10^{-7}$\,yr$^{-1}$ \citep{ckl+00}. However, the fact
that PSR~J1119$-$6127 glitched during less than 2\,yr of observations implies
that this pulsar likely glitches often and may prove to have a glitch activity
parameter on the order of that of the Crab pulsar, or larger. At the
opposite end of the spectrum lies the $\sim1700$ year old pulsar
PSR~B1509$-$58, which has not glitched in  21.3\,yr of
observations \citep{lkgm05}. 

The single observed glitch in \psr\ between 1996 and 2004 implies A$_g
\sim 0.002 \times 10^{-7}$\,yr$^{-1}$. Undetected glitches occuring in
the period prior to \rxte\ observations
would increase A$_g$ and could be the source of discrepant values of $n$ in
the literature. However, it is also possible that the pulsar did not glitch during
this time period, which would lower the overall value of A$_g$.
In this case, the range of reported values of $n$ would be due to timing
noise and/or pulse counting errors. The sparse nature of observations (and
in some cases data quality) between 1979 and 1996 precludes
conclusions on this point. Although the estimate of A$_g$ is based
on a single glitch and is therefore extremely sensitive to any future glitch
activity, it is clear that \psr\ glitches less than the Crab pulsar and
somewhat more than PSR~B1509$-$58. Thus, there is a range of possible glitch
activity amongst very young pulsars.

The variety of glitch activity is suggestive of a 
range of internal temperatures among young pulsars, since glitch activity is 
possibly associated with neutron-star temperature \citep{ml90}. In this 
theory, the very young pulsars such as the Crab, PSR~J1119$-$6127,
PSR~B1509$-$58 and \psr\ have small glitch activities
because they are relatively hot. Pulsars that are slightly older,
e.g. the ~10-kyr old Vela pulsar, are cooler and have larger glitch
activities, $\sim 8 \times 10^{-7}$\,yr$^{-1}$ \citep[][and references
therein]{dml02}. Finally, as pulsars become much older and cooler, they are believed to
eventually stop glitching entirely.  However, recent reports of glitches in
AXPs, which are thought to be hot neutron stars, argues that
temperature is not the only factor in determining glitch 
behaviour \citep{klc00,kg03,dos+03}. Regardless,
it is intriguing that there is such a wide variety of glitch
behavior in young pulsars.

\acknowledgments
This research made use of data obtained from the High
Energy Astrophysics Science Archive Research Center Online Service, provided
by the NASA-Goddard Space Flight Center. VMK is a Canada Research Chair and 
an NSERC Steacie Fellow. Funding for
this work was provided by NSERC Discovery Grant Rgpin 228738-03 and
Steacie Supplement Smfsu 268264-03.  Additional funding came from Fonds de
Recherche de la Nature et des Technologies du Quebec (NATEQ), the
Canadian Institute for Advanced Research, and the Canada Foundation
for Innovation.

\begin{center}
\begin{deluxetable}{lc}
\tablecaption{Parameters for \psr. \label{spin}}
\tablewidth{0pt}
\startdata
\hline
\multicolumn{2}{c}{Parameters for timing analysis} \\ \hline \hline
Dates (Modified Julian Day) &50150 -- 52935\\
Right Ascension\tablenotemark{a} (J2000) &$05^{\rm{h}}40^{\rm{m}} 11\fs16\pm0\fs04$\\
Declination\tablenotemark{a} (J2000)&$-69^{\circ} 19' 53\farcs9\pm0\farcs2$\\
\hline
\multicolumn{2}{c}{Parameters for partially coherent analysis} \\ \hline
\hline
Braking Index, $n$ &2.140(9)\\
Glitch Epoch (MJD) &51342(24) \\
$\Delta{\dot{\nu}}/{\dot{\nu}}$ ($10^{-4}$) &1.5(1) \\
\hline
\multicolumn{2}{c}{Parameters for phase-coherent analysis} \\ \hline \hline
Epoch (Modified Julian Day) &51197.0\\
$\nu$ (Hz)\tablenotemark{b} &19.80244383176(2)\\
$\dot{\nu}$ ($10^{-10}$~s$^{-2}$)\tablenotemark{b} &$-$1.878039597(8)\\
$\ddot{\nu}$ ($10^{-21}$~s$^{-3}$)\tablenotemark{b} &3.752027(2)\\
Braking Index, $n$\tablenotemark{c} &2.11(6)\\
Glitch Epoch (MJD) &51335(12)\\
$\Delta{\nu}/{\nu}$ ($10^{-9}$)\tablenotemark{b} &1.4(2)\\
$\Delta{\dot{\nu}}/{\dot{\nu}}$ ($10^{-4}$)\tablenotemark{b} &1.33(2)\\
\enddata
\tablenotetext{a}{Position fit from a grid search by minimizing the phase
residuals in a phase-coherent analysis, see \S \ref{sec:position}. }
\tablenotetext{b}{Quoted uncertainties are formal uncertainties as reported
by TEMPO.}
\tablenotetext{c}{Uncertainty determined from variation in $n$ as higher
order derivatives are fitted, as explained in \S \ref{sec:fully_coherent}.}
\end{deluxetable}
\end{center}

\normalsize
\clearpage

\normalsize
\clearpage

\normalsize
\clearpage

\normalsize
\clearpage

\normalsize
\clearpage

\normalsize
\clearpage

\end{document}